\documentclass{sig-alternate-2013}
\usepackage{color}
\usepackage{url}
\usepackage{tabularx}
\usepackage{xspace}
\usepackage{listings}
\usepackage{subfigure}
\usepackage{amssymb}
\usepackage{amsmath}
\usepackage{pifont}
\usepackage{graphicx}
\usepackage{times}
\usepackage{algorithm}
\usepackage{algorithmicx}
\usepackage[noend]{algpseudocode}
\usepackage{mathrsfs}
\usepackage{courier}
\usepackage{balance}
\usepackage{enumitem}

\newcommand{\xhdr}[1]{\vspace{2mm}\noindent{{\bf #1.}}}
\newcommand{\xhdrx}[1]{\vspace{1mm}\noindent{{\bf #1.}}}

\definecolor{orange}{rgb}{1,0.5,0}

\definecolor{red}{rgb}{1,0,0}

\newcommand{\hide}[1]{}

\newcommand{\hidex}[1]{}    
\newcommand{\denselist}{ \itemsep -2pt\topsep-10pt\partopsep-10pt }

\newcommand{\SNAP}{{\textsc{SNAP}}\xspace}

\algblock{ParFor}{EndParFor}
\algnewcommand\algorithmicparfor{\textbf{parfor}}
\algnewcommand\algorithmicpardo{\textbf{do}}
\algnewcommand\algorithmicendparfor{\textbf{parfor}}
\algrenewtext{ParFor}[1]{\algorithmicparfor\ #1\ \algorithmicpardo}
\algtext*{EndParFor}

\newfont{\mycrnotice}{ptmr8t at 7pt}
\newfont{\myconfname}{ptmri8t at 7pt}
\let\crnotice\mycrnotice%
\let\confname\myconfname%
\permission{Permission to make digital or hard copies of all or part of this work for personal or classroom use is granted without fee provided that copies are not made or distributed for profit or commercial advantage and that copies bear this notice and the full citation on the first page. Copyrights for components of this work owned by others than ACM must be honored. Abstracting with credit is permitted. To copy otherwise, or republish, to post on servers or to redistribute to lists, requires prior specific permission and/or a fee. Request permissions from permissions@acm.org.}
\conferenceinfo{SIGMOD'15,}{May 31--June 4, 2015, Melbourne, Victoria, Australia.}
\copyrightetc{Copyright \copyright~2015 ACM \the\acmcopyr}
\crdata{978-1-4503-2758-9/15/05\ ...\$15.00.\\
http://dx.doi.org/10.1145/2723372.2735369}

\begin{document}

\title{Ringo: Interactive Graph Analytics\\
on Big-Memory Machines}
%
%
%
%
%

\numberofauthors{1} 
%
\author{
%
%
\alignauthor
Yonathan Perez, Rok Sosi\v{c}, Arijit Banerjee, Rohan Puttagunta, Martin Raison,\\ Pararth Shah, Jure Leskovec\\
	\affaddr{Stanford University}\\
        \email{\{yperez, rok, arijitb, rohanp, mraison, pararth, jure\}@cs.stanford.edu}
}

\maketitle

\begin{abstract}

We present Ringo, a system for analysis of large graphs. Graphs provide a way to represent and analyze systems of interacting objects (people, proteins, webpages) with edges between the objects denoting interactions (friendships, physical interactions, links). Mining graphs provides valuable insights about individual objects as well as the relationships among them. 

In building Ringo, we take advantage of the fact that machines with large memory and many cores are widely available and also relatively affordable. This allows us to build an easy-to-use interactive high-performance graph analytics system. Graphs also need to be built from input data, which often resides in the form of relational tables. Thus, Ringo provides rich functionality for manipulating raw input data tables into various kinds of graphs. Furthermore, Ringo also provides over 200 graph analytics functions that can then be applied to constructed graphs.

We show that a single big-memory machine provides a very attractive platform for performing analytics on all but the largest graphs as it offers excellent performance and ease of use as compared to alternative approaches. With Ringo, we also demonstrate how to integrate graph analytics with an iterative process of trial-and-error data exploration and rapid experimentation, common in data mining workloads.

\end{abstract}

\vspace{2mm}\noindent{\bf Categories and Subject Descriptors:} 
H.2.4 [{\bf Information Systems}]: Database Management---{\em Systems}

\noindent{\bf General Terms:} Algorithms, Performance




\section{Introduction}
\label{sec:intro}

Detecting expert users in a question-answering forum, tracing the propagation of information in a social network, or reconstructing the Internet network topology from a set of traceroutes are examples of tasks faced by today's data scientists. A common theme to all these examples is that they involve input data manipulation as well as graph analytics, where graphs are analyzed using various graph algorithms. To solve such problems and extract valuable insights, data scientists must be able to quickly construct graphs from input data, and analyze the graphs using various graph algorithms. 

Thus, in order to support the work of data scientists, one requires a system that offers a rich set of {\em graph manipulation and analysis algorithms}. 
As graphs are rarely given as input, but have to be {\em constructed} from input data, a modern graph analytics system has to support easy manipulation of input data in order to build a desired graph. 
Arguably the most common form of input data are {\em relational tables} and while in principle tables can be used to represent graphs, dedicated graph structures where the neighbors of each node are easily accessible are more efficient for most graph computations.
Thus, the system also needs to provide a way to {\em convert} between graphical and tabular data structures and data representations. 

And last, the system also needs to provide {\em fast execution times} suitable for {\em interactive} use.

In summary, the desiderata for a modern data science oriented graph analytics system are:
\begin{enumerate}
\denselist
\item[(1)] Ability to process \hide{richly annotated} large graphs, on the order of hundreds of millions of nodes and billions of edges,
\item[(2)] Fast execution times that allow for interactive, exploratory use (as opposed to batch-mode use),
\item[(3)] Easy to use front-end that provides many graph algorithms in a commonly used high-level programming language,
\item[(4)] Large number of efficient ready-to-use graph algorithms,
\item[(5)] Rich support for transformations of input data to graphs.
\end{enumerate}

\vspace{-1mm}
There are many challenges in building such systems. For example, what underlying hardware infrastructure shall one use? A cluster or a big server? How does one design data structures for tables and graphs that are efficient, flexible and fast? What operations are needed for building graphs from input data tables? What are the considerations for end-to-end graph analytics systems?

\xhdr{Ringo: Graph analytics on a big-memory machine}
We present Ringo, an in-memory interactive graph analytics system that scales to large graphs. Ringo combines an easy-to-use Python front-end and a scalable parallel C++ back-end, which is responsible for rapid data handling and manipulation. Ringo provides functionality for efficiently building graphs from input data tables, for converting the tables to an efficient graph data structure, and for analyzing graphs using over 200 different graph functions through its core graph analytics package SNAP\footnote{SNAP is currently downloaded about a thousand times per month and actively used in our research group, as well as by over 500 students in Stanford University courses.}. Ringo source code is open\footnote{\url{http://snap.stanford.edu/ringo}}. 

\begin {figure}[t]
\centering
\includegraphics[width=0.9\columnwidth]{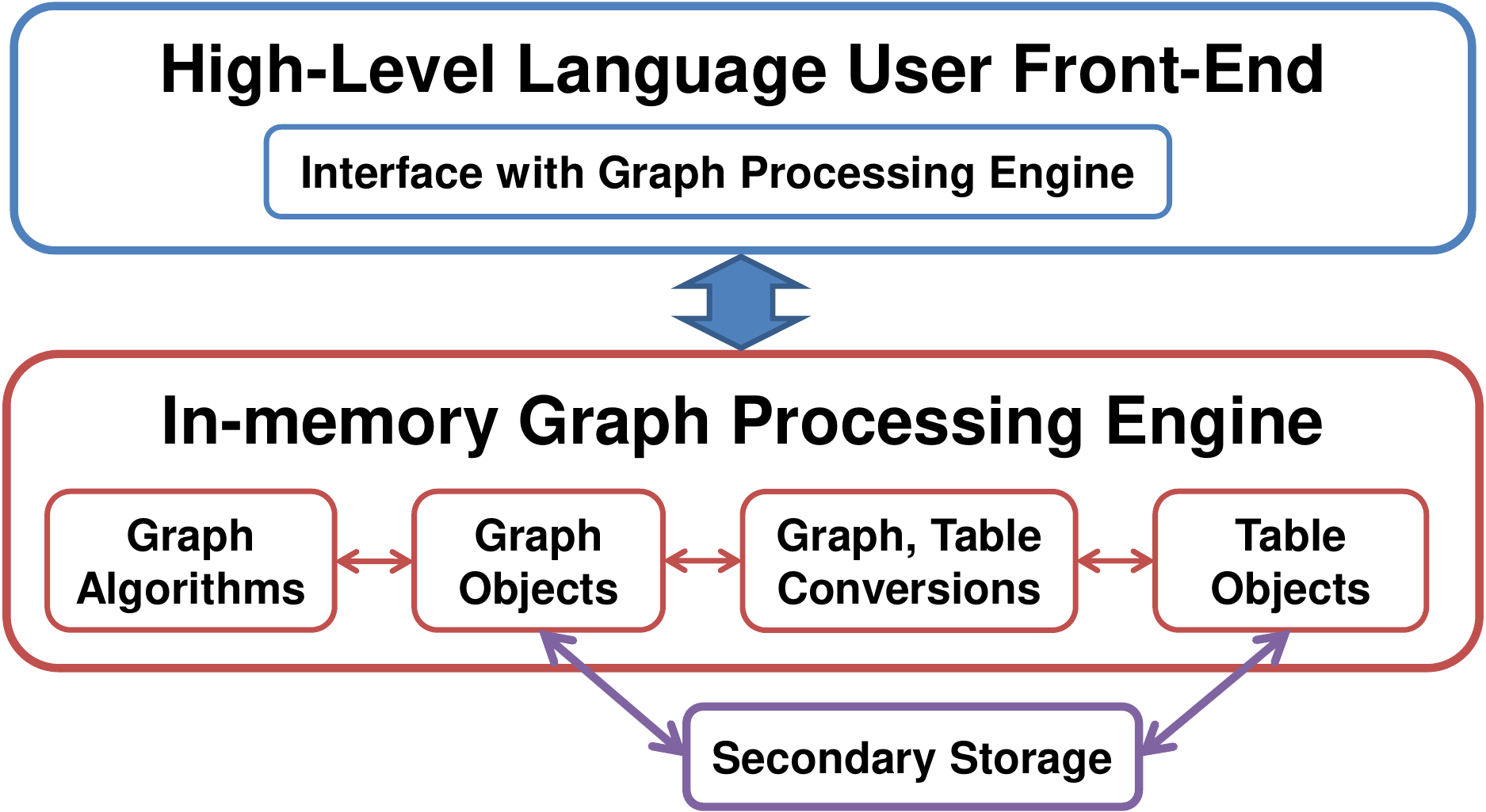}
\vspace{-2mm}
\caption{Ringo system overview.}
\label{fig:ringo}
\vspace{-3mm}
\end{figure}

Recent research in graph analytics systems has been focused on distributed computing environments~\cite{PowerGraph,gonzalez2014graphx,han2014experimental,Pregel,qin2014scalable,satish2014,shao2013trinity} or single-machine systems utilizing secondary storage~\cite{TurboGraph,kim2014opt,GraphChi}.
Such systems offer scalability in the number of cores or in available throughput and size of the secondary storage, but these benefits come at a high price of increased communication cost, increased system complexity, and challenges when programming non-trivial graph algorithms. On the other hand, big-memory, multi-core machines are becoming affordable and widely available; a machine with 1TB of main memory and 80 cores costs around \$35K.

We observe that most graphs being analyzed today comfortably fit in the memory of one such ``big-memory'' machine. Graph based computations require random data access patterns and exhibit notoriously poor data locality, so a single big-memory machine seems a natural hardware choice for analytics of all-but-largest graphs.

Ringo is built on the assumption that graphs being analyzed fit in memory of a single machine. This approach has significant benefits in that there is no network communication overhead, no need for managing the secondary storage, and that the programming model and the system use are straightforward. Even though the raw input data might not fit into the main memory initially, data cleaning and manipulation often result in significant data size reduction, so that the ``interesting'' part of the data nicely fits into the main memory. 

Ringo showcases that a single multi-core machine offers a suitable platform for interactive graph analytics, while matching the performance of the fastest distributed graph processing systems (Section~\ref{sec:experiments}).
Figure~\ref{fig:ringo} illustrates Ringo.

The key features of Ringo are as follows:
\begin{itemize}[noitemsep]
\item A system for interactive and exploratory analysis of large graphs with hundreds of millions or even billions of edges,
\item Tight integration between graph and table processing and efficient conversions between graphs and tables,
\item Powerful operations to construct various types of graphs,
\item Ringo runs on a single machine with a large main memory, simplifying programming significantly and out performing distributed systems on all but the largest graphs.
\end{itemize}

The rest of the paper is organized as follows.
Section~\ref{sec:system} provides a system overview.
System evaluation and benchmarks are presented in Section~\ref{sec:experiments}. 
Usage scenarios are described in Section~\ref{sec:demos}. 
Finally, we conclude in Section~\ref{sec:conclusion}.

\hidex{{\color{orange}
\section{Related Work}
\label{sec:related}
\input{p080related}
}}

\vspace{-2mm}
\section{System Overview}
\label{sec:system}

\vspace{-2mm}
\subsection{Design Choices}

In designing Ringo for interactive graph analytics, we were informed by two insights. The first insight is that all but the largest graphs being analyzed today fit comfortably in memory of a big-memory machine. The second insight is that graphs are not analyzed in isolation, but are part of a larger data mining workflow, which requires that, in addition to graph operations, table analytics operations are integrated in the system as well.
We discuss those insights and design choices in more detail next.

\xhdrx{Graph sizes}
Our goal is to provide fast execution time for commonly analyzed, average size graphs rather than provide a scalable solution for extremely large (and rare) graphs. 

Classifying graphs from a widely used repository~\cite{snapnets} of publicly available real-world networks (covering a wide range of areas, such as the Internet, social networks, road networks, Web networks) according to the number of edges shows that 90\% out of 71 graphs have less than 100 million edges and only one graph has more than 1 billion edges (Table~\ref{table:snap_size}). Most graphs fit in a few GB of RAM, and the largest graph requires only about 30GB of RAM.
Furthermore, very few real-world, non-synthetic graphs with more than one billion edges are discussed regularly in present research literature, Twitter2010 with 1.5B edges and Yahoo Web graph with 6.6B edges being the most common.  Assuming 20 bytes of storage per edge, even the Yahoo Web graph requires only about 135GB and fits easily in a 1TB RAM machine. Other studies confirm our observation that most analytics datasets are limited in size~\cite{appuswamy2013scale,rowstron2012nobody}.

\begin{table}
	\centering
	\begin{tabular}{c c}
	\hline
	Number of Edges & Number of Graphs \\
	\hline
	\textless 0.1M &  16 \\
	0.1M -- 1M &  25 \\
	1M -- 10M &  17 \\
	10M -- 100M &   7 \\
	100M -- 1B &   5 \\
	\textgreater 1B &   1 \\
	\hline
	\end{tabular}
	\vspace{-2mm}
	\caption{Graph size statistics of 71 graphs publicly available in the
	Stanford Large Network Collection. 90\% of graphs have less than 100M edges. Only one graph has more than 1B edges.}
	\label{table:snap_size}
	\vspace{-2mm}
\end{table}

Based on this evidence we conclude that big-memory machines allow for storing and processing large graphs in-memory. As we show later in the experimental section, such machines also have sufficient compute power to execute graph analytics at speeds comparable to the fastest distributed graph processing systems.

\begin {figure}[t]
	\centering
	\includegraphics[width=0.95\columnwidth]{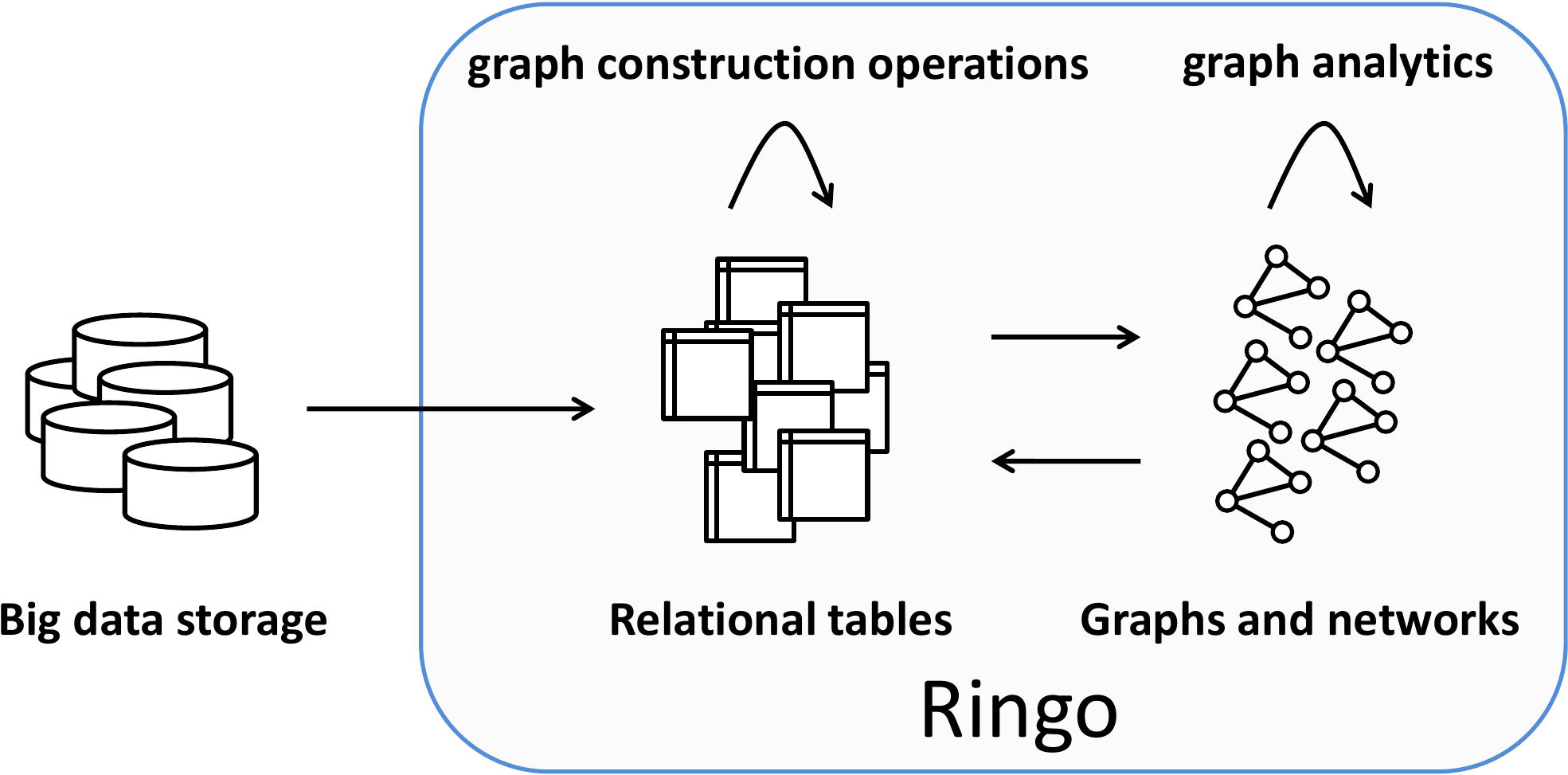}
	\vspace{-3mm}
	\caption{Graph analytics workflow. Using a system like Hadoop, data of interest is extracted from a big data storage into a set of relational tables. Graphs are built using graph construction operations on tables; the results are converted to graphs. Graph analytic operations are then applied to the graphs. Results of graph operations are added back to tables.}
	\label{figure:workflow}
	\vspace{-6mm}
\end{figure}

\xhdrx{Graph analytics workflow}
Graph analytics often follow the workflow presented in Figure~\ref{figure:workflow}. 
Raw data is stored in a big data repository and handled by a system like Hadoop to extract the initial data for analysis.
The extracted data is organized in a set of relational tables and a major part of the graph analytics workflow is then to construct many different graphs from the tables.
Once graphs are built, we require graph specific operations, such as PageRank, connected components, shortest paths. Although graphs can be represented as relational tables and graph operations can be implemented using relational operations, we find it is more efficient (in terms of memory as well as speed) to have optimized graph-specific data structures.
Thus, given graphs represented as tables, the next step is to convert these tables to a graph representation. We then execute graph operations and integrate the results back to tables. The result is an iterative process, where data can be rapidly converted from tables to graphs and vice versa.

To provide fast execution speed in Ringo, we implemented table processing as part of the system, which allows us to tightly integrate table and graph processing and to rapidly convert large datasets from one representation to another.

\vspace{-1mm}
\subsection{Graphs in Ringo}

\xhdrx{Ringo graph representation}
A critical operation for fast execution of graph algorithms is quick access to neighboring nodes and edges. Additionally, we require that our graph representation is dynamic, so in existing graphs nodes and edges can be quickly added or removed.
A challenge is thus to strike a balance between opposing requirements for fast access to a node's neighborhood and having a dynamic graph structure.

One possible approach for efficient graph representation would be to use the Compressed Sparse Row format~\cite{CSR}. This format uses two vectors, a vector for nodes and a vector for edges. The edge vector is sorted by source nodes and indexed by the node vector for fast access. While this approach offers high performance for traversal operations, it does not perform well for dynamic graphs, since graph updates cause prohibitive maintenance costs of the single big edge vector (e.g., deleting a single edge requires time linear in the total number of edges in the graph).

Ringo supports dynamic graphs by representing a graph as a hash table of nodes. Each node maintains sorted adjacency vector of neighboring nodes. Space requirements for this representation are similar to those of the Compressed Sparse Row format. However, we found that representing graphs as a hash table of nodes and the associated adjacency vectors does not dramatically impact the performance of graph algorithms (e.g., deleting a single edge only requires time linear in the node degree).

\xhdrx{Ringo graph operations}
To provide a rich set of graph constructs and algorithms, Ringo builds on the publicly available Stanford Network Analysis Platform (\SNAP)~\cite{snap}. SNAP provides more than two hundred out-of-the-box graph constructs and algorithms that are available for use on Ringo in-memory graph data structure.

\vspace{-1mm}
\subsection{Tables in Ringo}
\label{sec:tables}

\xhdrx{Table representation}
In addition to graph objects, Ringo implements its own native relational table objects to allow for efficient and flexible parallel implementations of 
operations important for graph construction, to support fast conversions into graph objects, and to avoid any performance overheads
related to frequent transitions to and from external systems during the iterative data analysis process.
Tables in Ringo have a schema, which defines table columns and their types (integer, floating point, or string).
Since tables have been studied extensively~\cite{databases}, we only describe some Ringo specific details here.

As most tabular operations in Ringo are graph related and primarily use iterations over columns, Ringo table representation optimizes this use by implementing tables with a column based store.
In Ringo each row has a persistent unique identifier. This allows for fast in-place grouping, filtering and selection. Moreover, identifiers allow for fine-grained data tracking, so the user can identify data records even after they undergo a complex set of operations.

Ringo provides basic relational operations on table objects, such as select, join, project, group \& aggregate, set operations, order (sort), and similar.
In addition, Ringo also provides a number of advanced graph construction operations, described next.

\xhdrx{Graph construction}
In order to construct a graph, we first manipulate input data tables into an {\em edge table} that has two columns, a column with edge source nodes and a column with edge destination nodes. Once the edge table is constructed we transform it into a Ringo graph in-memory data structure.

In some cases, the edge table can be constructed using basic relational operations, such as join and select.
However, often graph construction requires advanced operations unique to Ringo. Ringo allows for creating edges based on node similarity or temporal order of nodes. Ringo implements \emph{SimJoin}, which joins two records if their distance is smaller than a given threshold, and \emph{NextK}, which joins predecessor-successor records.

\vspace{-1mm}
\subsection{Converting Between Tables and Graphs}
\label{sec:conversion}

Fast conversions between graph and table objects are essential for data exploration tasks involving graphs.
Without the loss of generality we limit our discussion to conversions of directed graphs. 

\xhdrx{Conversion of tables to graphs}
The goal is to generate a directed graph $G = (V,E)$ from table $T$, with edge source column $S$ and edge destination column $D$.
Nodes $V$ in the graph are defined by unique values in columns $S$ and $D$.
And each row $r \in T$ defines an edge $e \in E$ with the source node provided in column $S$ and the destination node provided in column $D$, $e = (r_S, r_D)$.

A directed graph in Ringo is represented as a node hash table, where each node contains two sorted adjacency vectors providing its in-neighbors and out-neighbors. The problem of converting a table to a graph is how to transform an edge list given by two columns in a table to a node hash table with sorted neighbor vectors.
The challenge is to transform tables with hundreds of millions of rows at speeds that make the system suitable for interactive use.

We experimented with several approaches and found that a ``sort-first'' algorithm works the best. 
The algorithm builds a graph representation from a table by first making copies of the source and destinations columns, then sorting the column copies, computing the number of neighbors for each node, and then copying the neighbor vectors to the graph hash table.
Advantages of our method are that sorting can be done in parallel and that it does not require any thread-safe operations on vectors and hash tables. While concurrent access is still performed, there is no contention among the threads, which minimizes locking and allows fast execution on multi-core machines. Since the number of neighbors is calculated explicitly, there is also no need to estimate the size of the hash table or neighbor vectors in advance.

\xhdrx{Conversion of graphs to tables}
The conversion from a graph to a table involves building a node table  or an edge table. This conversion can be easily preformed in parallel by partitioning the graph's nodes or edges among worker threads, pre-allocating the output table, and assigning a corresponding partition in the output table to each thread. The threads iterate over nodes or edges in their graph partitions and write the output to their assigned partitions in the output table.

\vspace{-1mm}
\subsection{Ringo Implementation}

We highlight aspects of Ringo implementation that allow for high-performance graph processing in an interactive environment.

\xhdrx{High-level language front-end}
The user interacts with Ringo through a Python module. 
Ringo front-end utilizes our graph processing engine to execute time critical parts. We use SWIG~\cite{swig} to connect the Python front-end with the parallel C++ table and graph processing back-end engine.

\xhdrx{High-performance graph processing engine}
Ringo graph processing engine is based on \SNAP~\cite{snap}, a highly efficient C++ graph analysis library that implements complex graph classes and a rich set of graph constructs and algorithms.
For Ringo, we have expanded \SNAP with several new components, including table processing, conversions between tables and graphs, and parallel graph algorithms. OpenMP was used to parallelize critical loops in the code for full utilization of our target multi-core platforms.

\xhdrx{Concurrent hash tables and vectors}
The OpenMP layer relies on fast, thread-safe operations on concurrent hash tables and vectors, which are critical for achieving high performance of graph operations.  We implemented an open addressing hash table with linear probing~\cite{lang2013massively}. 
To support fast graph construction, we extended the node hash table with thread-safe insertions to a node's adjacency vector. Concurrent insertions to a vector are implemented by using an atomic increment instruction to claim an index of a cell to which a new value is inserted.

\vspace{-3mm}
\section{System Performance}
\label{sec:experiments}

\vspace{-1mm}
In this section, we show Ringo performance on a single, big-memory, multi-core machine.
Experiments demonstrate that such machines are able to execute critical operations at speeds that are needed for interactive graph analysis.

\xhdrx{Experimental datasets and setup}
\begin{table}[t]
\centering
\begin{tabular}{l r r}
\hline
Graph Name & LiveJournal~\cite{LiveJournal} & Twitter2010~\cite{Twitter2010} \\
\hline
Nodes & 4.8M & 42M \\
Edges & 69M & 1.5B \\
Text File Size & 1.1GB & 26.2GB \\
In-memory Graph Size & 0.7GB & 13.2GB \\
In-memory Table Size & 1.1GB & 23.5GB \\
\hline
\end{tabular}
\caption{Experiment graphs. {\em Text Size} is the size of the input text file, {\em Graph Size} is the size of the corresponding Ringo graph object, and {\em Table Size} is the size of the Ringo table object.}
\label{table:graphs}
\end{table}
For our benchmarks, we use two popular graphs, {\it LiveJournal} and the larger {\it Twitter2010} (Table~\ref{table:graphs}), that have been widely used for benchmarks of other large-scale graph processing systems~\cite{Patric,PowerGraph,TurboGraph,hong2012green,GraphChi,GraphLab,Pregel}.  In many cases, graphs used for analyses are of sizes similar to {\it LiveJournal}.
However, larger graphs do exist and can be easily processed on machines with 1TB RAM.
For example, the {\em Twitter2010} graph takes only about 13GB of main memory in Ringo (Table~\ref{table:graphs}), which means that the system could easily process graphs that are an order of magnitude larger.

Our measurements were performed on a machine with 1TB RAM and 4x Intel CPU E7-4870 at 2.40GHz, each CPU has 10 cores and supports 20 hyperthreads for a total of 80 hyperthreads, running CentOS 6.4.  A similar machine costs \$35K as of Nov 2014.

\begin{table}[t]
	\centering
	\begin{tabular}{ l r r }
	\hline
	Operation & LiveJournal & Twitter2010 \\
	\hline
	PageRank & 2.76s & 60.5s \\
	Triangle Counting & 6.13s & 263.6s \\
	\hline
	\end{tabular}
	\vspace{-1mm}
	\caption{Performance of parallel graph algorithms for PageRank and Triangle Counting on a single big-memory machine with 80 cores. For PageRank, ten iterations were timed.}
	\label{table:paralg}
	\vspace{-4mm}
\end{table}

\xhdrx{Parallel graph algorithms}
We demonstrate graph analysis capabilities of big-memory machines by using PageRank~\cite{PageRank} and undirected triangle counting---two key graph algorithms that are often used for benchmarking purposes. PageRank is an example of an iterative message-passing-like computation. We measured the runtime of 10 iterations. Triangle counting is directly related to relational joins, one of the central problems in database systems~\cite{NgoRR13}.
We ran each experiment 5 times, and report the average runtimes. Table~\ref{table:paralg} shows the results. It is worth comparing the performance of Ringo to the recently published results on the {\it Twitter2010} graph.

For triangle counting, a recently published method~\cite{kim2014opt} took 469s using 1 machine with 6 cores and SSD based secondary storage, while a different, distributed approach took 564s on a cluster of 200 processors~\cite{Patric}. Ringo takes 263s on 1 machine with 80 hyperthreads. The implementation of the algorithm in Ringo was simple, using a straightforward approach, similar to~\cite{Patric} and parallelizing the execution with a few OpenMP statements.

PowerGraph~\cite{PowerGraph} is one of the highest performing graph processing systems. Using 64 machines with 512 cores, PowerGraph took 3.6s per iteration of the PageRank algorithm. On the other hand, Ringo takes around 6s per iteration, while using 6x fewer cores, only one machine and 13GB RAM. Similarly to the implementation of triangle counting, PageRank implementation in Ringo is based on a straightforward, sequential algorithm with a few OpenMP statements for parallel execution.  Even though the purpose of Ringo is not to be the fastest graph engine, comparisons show the viability of big-memory machines for processing all-but-the-largest graphs.

In addition to fast execution times, Ringo graph processing also keeps a low memory footprint. The computation of 10 iterations of PageRank on the {\it Twitter2010} graph had a memory footprint of 18.3GB, and triangle counting on that same graph had a memory footprint of 22.6GB. In both cases the memory footprint was less than twice the size of the graph object itself.

\xhdr{Table operations}
\begin{table}[t]
\centering
\begin{tabular}{ l r r }
\hline
Dataset & LiveJournal & Twitter2010 \\
\hline
Select 10K, in place & \textless 0.2s & 1.6s \\
Rows/s & 405.9M & 935.3M \\
\hline
Select all-10K, in place & \textless 0.1s & 1.6s \\
Rows/s & 575.0M & 917.7M \\
\hline
Join 10K & 0.6s &4.2s \\
Rows/s & 109.5M & 348.8M \\
\hline
Join all-10K & 3.1s & 29.7s \\
Rows/s & 44.5M & 98.8M \\
\hline
\end{tabular}
\caption{Ringo performance of Select and Join operations on tables. Numbers give measured times on our test datasets. The {\em Rows/s} gives the processing rate in millions of table rows processed per second. 
The processing rate for {\em Join} takes the sizes of both input tables into account.
}
\label{table:table_ringo}
\vspace{-5mm}
\end{table}
While the focus of Ringo is on graph analytics, table operations are necessary for processing the data prior to graph construction and are a fundamental part of the analysis process. To support interactive data exploration the table operations have to execute quickly and at speeds comparable to graph operations.

We benchmark Ringo on two essential table operations, \emph{select} and \emph{join} (Table~\ref{table:table_ringo}).
For select benchmarks, rows are chosen based on a comparison with a constant value.
The value is determined so that it either selects 10,000 elements from the table or all elements except 10,000.
The purpose of two measurements is to show performance when the output is either very small or similar in size to the input table.
We show results for the select in-place operation, where the current table is modified.

For join benchmarks, the input table is joined with a second, single column table.
The values in the second table are chosen so that the output table has either 10,000 elements or all elements from the input table except 10,000. Ringo join operation always produces a new table object. Overall, results in Table~\ref{table:table_ringo} demonstrate that Ringo offers robust performance over a range of scenarios.

\xhdrx{Conversions between tables and graphs}
\label{exp:conversions}
Next we present performance of Ringo algorithms when converting tables to graphs and vice versa (discussed in Section~\ref{sec:conversion}).

\begin{table}[t]
\centering
\begin{tabular}{ l r r }
\hline
Graph Name & LiveJournal & Twitter2010 \\
\hline
Table to graph & 8.5s & 81.0s \\
Edges/s & 13.0M & 18.0M \\
\hline
Graph to table & 1.5s & 29.2s \\
Edges/s & 46.0M & 50.4M \\
\hline
\end{tabular}
\vspace{-2mm}
\caption{Execution times for converting tables to graphs and vice versa. The {\em Edges/s} row gives the processing rate in millions of edges processed per second.}
\label{table:conversion_ringo}
\vspace{-2mm}
\end{table}

Table~\ref{table:conversion_ringo} gives Ringo execution times for conversions of {\it LiveJournal} and {\it Twitter2010} datasets between table and graph representations.
For example, in the specific case of the {\it Twitter2010} graph, the conversion of the table containing a single large edge vector to a graph means that a table with 1.5B rows must be traversed, and 1.5B pairs of node identifiers are extracted. Each element of a pair is inserted to the node hash table in the graph and elements are added to the two corresponding adjacency vectors, which must be sorted. This results in a graph representation with a total of 42M nodes, and 84M vectors containing adjacent nodes.

Overall, the conversion from a table to a graph is performed at a rate of over 10M table rows or graph edges per second and about 50M edges per second in the opposite direction, so graphs with tens of millions of edges can be processed in seconds. The conversion scales well as the processing rate does not degrade for large graphs.

\xhdrx{Sequential performance}
For moderately large real-world graphs such as {\it LiveJournal}, even sequential implementations of graph algorithms are fast enough for interactive analysis.
We measured the runtime of sequential implementations of 3 commonly used graph algorithms on the {\it LiveJournal} graph: \emph{3-core} of the graph, \emph{single source shortest path} (runtime averaged over 10 random sources), and finding \emph{strongly connected components} (Table~\ref{table:SequentialAlgs}). All algorithms executed in about 30 seconds or less. For larger graphs, parallel implementations of graph algorithms are needed, and we are currently expanding the set of parallel algorithm implementations available through Ringo.

\begin{table}[t]
	\centering
	\begin{tabular}{ l r}
	\hline
	Algorithm & Runtime \\
	\hline
	3-core & 31.0s \\
	\hline
	SSSP & 7.4s\\
	\hline
	SCC & 18.0s\\ \hline
	\end{tabular}
	\caption{Runtime of single-threaded implementations of commonly used graph algorithms on the {\em LiveJournal} graph: 3-core, single source shortest path (SSSP), and strongly connected component decomposition (SCC).}
	\label{table:SequentialAlgs}
	\vspace{-3mm}
\end{table}

\section{Ringo Scenarios}
\label{sec:demos}

We demonstrate Ringo on real-world graph analytic scenarios, illustrating its applicability, ease of use, and performance.

\subsection{End-to-end Graph Analytics}

Our demo will showcase Ringo capabilities in an end-to-end graph analytics scenario. SIGMOD attendees will be able to interact with Ringo and observe ease of use and integration of table and graph operations.

\xhdrx{Finding Java experts on StackOverflow}
We will show a realistic and representative use case where the goal is to identify top Java experts in the StackOverflow user community. For demonstration we will use complete data from StackOverflow, which is the world's largest question-answering website, where users post questions, then others answer them.  As the answers are given, the person posting the question has the option of picking the best answer by ``accepting'' it.
In order to identify top experts our demo will start with complete StackOverflow data (8M questions, 14M answers, 34M comments). The demo will then follow these steps: manipulate tables to build a graph which connects users providing Java related questions and answers, and use a graph algorithm to identify top Java experts. 

\xhdr{Demonstrantion scenario}
The SIGMOD attendee will first load complete StackOverflow data\footnote{Freely available at \url{http://data.stackexchange.com}.} in a form of relational tables. The attendee will then manipulate the input tables in order to build a graph representing the interactions in the forum's social network. For example, one way to build a graph is to connect users who answered the same question. A different way is to connect StackOverflow users that answered each other's questions. Ringo provides rich functionality for the SIGMOD attendee to build various kinds of graphs based on the StackOverflow data. Once the graph is built, the attendee will identify important nodes in the graph. Ringo implements over 200 different graph analytics algorithms (e.g., PageRank, Hits, and various other node centrality measures) that the attendee may try out to find Java experts.

Below, we show Ringo Python commands for the above demo\footnote{For complete code for the demo, see \url{http://snap.stanford.edu/ringo}}. The attendee will load the StackOverflow posts table {\tt P}, extract all the Java posts {\tt JP}, and build two new tables: questions table {\tt Q} and answers table {\tt A}:
\begin{lstlisting}
 P = ringo.LoadTableTSV(schema,'posts.tsv')
 JP = ringo.Select(P,'Tag=Java')
 Q = ringo.Select(JP,'Type=question')
 A = ringo.Select(JP,'Type=answer')
\end{lstlisting}
Next, the attendee will build a graph and apply the PageRank algorithm on it. Ringo primitives for this task are:
\begin{lstlisting}
 QA = ringoJoin(Q,A,'AnswerId','PostId')
 G = ringo.ToGraph(QA,'UserId-1','UserId-2')
 PR = ringo.GetPageRank(G)
 S = ringo.TableFromHashMap(PR,'User','Scr')
\end{lstlisting}
The {\em Answers} table on column {\em PostId} is joined with the {\em Questions} table on column {\em AnswerId}.
The resulting {\em QA} table has two {\em UserId} columns, corresponding to the users that asked questions and the users whose answers were accepted. 
Using these user columns in {\em QA}, {\tt ToGraph()} transforms the {\em QA} table into an optimized graph data structure {\em G}, where each node represents a user and an edge indicates that an answer by the destination node was accepted by the source node.
{\tt GetPageRank()} calculates the PageRank scores of all the nodes in {\em G}.
The remaining line builds the final table {\em S} with the users' PageRank scores.

The above example clearly demonstrates that a system cannot treat graph analytics in isolation, but that graph analytics needs to be integrated with table operations. 
With Ringo we will demonstrate a system, where graph and table operations are tightly integrated. Besides implementing table operations as part of the system, in Ringo this integration involves fast conversions between table and graph data structures as well as customized table operations, suitable for graph analytics. 

\xhdrx{Ringo ease of use}
We will also demonstrate several factors that contribute to ease of use of Ringo: integrated processing of graphs and tables, Python front-end, and execution on a single machine. 

\xhdrx{What the attendees will see and do}
The attendees will be able to load the StackOverflow dataset in Ringo and perform graph analytics on the dataset. First, an attendee will be able to execute operations that identify the top Java experts on StackOverflow.
Second, the attendee will be able to vary the parameters to identify top experts on StackOverflow in other programming languages and topics of interest, or change the operations to explore alternative measures of expertise.
Third, the attendee will be invited to an open exploration of the StackOverflow dataset by applying Ringo graph analytics capabilities.

\vspace{-1mm}
\subsection{Performance Demonstration}

This part of our demo will center around Ringo performance. Attendees will be able to observe system performance and resource utilization on a range of input datasets and hardware platforms.

\xhdrx{Graph analytics operations}
Several types of operations are critical for graph analytics: graph operations, table operations, conversions between tables and graphs, and data input/output. 
For the demo, we will show performance of a range of Ringo operations on the LiveJournal and the Twitter2010 datasets. LiveJournal is a typical representative of a graph dataset, while the Twitter2010 dataset is one of the largest publicly available real-world graph datasets.

\xhdrx{Hardware used for demonstration}
Big-memory machines provide a viable platform for interactive graph analytics. Large RAM on such machines can deliver fast random access, required by graph algorithms.
In our demo, we will show how Ringo performs on a wide range of operations on a big-memory machine. For our demo, we will use a machine with 1TB RAM and 4x Intel CPU E7-4870 at 2.40GHz, running CentOS 6.5. Each CPU has 10 cores and supports 20 hyperthreads for a total of 80 hyperthreads.

Ringo also works well on standard personal computers, provided that the datasets fit in the RAM available. As discussed before, 90\% of graphs in SNAP require only a few GB of RAM (see Table~\ref{table:snap_size}), so they can fit in memory of a typical desktop or laptop.
We will demo Ringo also on a laptop with 8GB RAM and an Intel Core i5-4258U CPU at 2.40GHz with 2 cores, costing around \$1,500.

\xhdrx{What the attendees will see and do}
The attendees will be able to use Ringo on a big-memory machine and on a personal computer.
Initially, the attendee will choose either a big-memory machine or a laptop environment and load one of the datasets, LiveJournal or Twitter2010. (Only LiveJournal will be available on the laptop due to RAM limitations.) Next, the attendees will be able to execute a wide range of graph algorithms and observe their performance and the hardware utilization.

\vspace{-2mm}
\section{Conclusion}
\label{sec:conclusion}

\vspace{-1mm}
We presented Ringo---a system for interactive and exploratory analysis of large graphs with hundreds of millions of nodes and edges.
Ringo exposes its functionality through high-level Python programming language, the language of choice of today's data scientists.
Tight integration between graph and table processing and efficient conversions between graphs and tables allow for powerful operations that make it easy to perform complex analytic tasks.

Overall, we demonstrate that big-memory machines are suitable for performing interactive analyses of all, but the very largest graphs. In many real-world scenarios graph sizes are well below a terabyte and in such cases big-memory machines have significant benefit over large distributed clusters. Single machines are easier and more efficient to program, while the cost and complexity of cross-machine communication and scheduling are eliminated.

\xhdrx{Acknowledgments}
\label{sec:acknowledgements}
We thank Vikesh Khanna, Chantat Eksombatchai, Sheila Ramaswamy, Nicholas Shelly, Nikhil Khadke, and Jason Jong for their assistance with Ringo implementation, and Andrej Krevl for his help with the experimental setup.
%
Yonathan Perez is supported by P. Michael Farmwald Stanford Graduate Fellowship. 
This research has been supported in part by NSF
IIS-1016909,              
IIS-1149837,       
IIS-1159679,              
CNS-1010921,              
DARPA XDATA,            
DARPA GRAPHS,           
DARPA SIMPLEX,
Boeing,                    %
Facebook,
Volkswagen,                 
Yahoo, and
SDSI.
\vspace{-2mm}
%
\bibliographystyle{abbrv}

%
%


\end{document}